\documentclass[useAMS,usenatbib,usegraphicx]{mn2e}

\usepackage{graphicx}
\usepackage[latin1]{inputenc}
\usepackage{color}
\usepackage{times}
\usepackage{natbib}
\usepackage{setspace}
\newif\ifAMStwofonts
\AMStwofontstrue
\definecolor{red}{rgb}{1,0.,0.}

\newcommand{\morgana}{{\sc morgana}}
\newcommand{\bbrel}{$M_{\rm bh}$-$M_{\rm sph}$ relation}
\newcommand{\msun}{{\rm M}_\odot}

\def\lesssim{\lower.5ex\hbox{$\; \buildrel < \over \sim \;$}}
\def\gtrsim{\lower.5ex\hbox{$\; \buildrel > \over \sim \;$}}
\title[Breaking the BH-Bulge relation] {Interpreting the possible
  break in the Black Hole - Bulge mass relation}

\author[Fontanot, Monaco \& Shankar]{
  \parbox[t]{\textwidth}{Fabio Fontanot$^1$\thanks{E-mail:
      fontanot@oats.inaf.it}, Pierluigi Monaco$^{2,1}$, Francesco Shankar$^3$}
    \vspace*{8pt}\\
    $^1$ INAF - Astronomical Observatory of Trieste, via G.B. Tiepolo 11, I-34143 Trieste, Italy \\
    $^2$ Dipartimento di Fisica, Sezione di Astronomia, via G.B. Tiepolo 11, I-34143 Trieste, Italy \\
    $^3$ Department of Physics and Astronomy, University of Southampton, Southampton SO17 1BJ, UK \\
}

\begin{document}
\date{Accepted ... Received ...}

\maketitle

\begin{abstract}
Recent inspections of local available data suggest that the almost
linear relation between the stellar mass of spheroids ($M_{\rm sph}$)
and the mass of the super massive Black Holes (BHs) residing at their
centres, shows a break below $M_{\rm sph} \sim 10^{10}\ \msun$, with a
steeper, about quadratic relation at smaller masses.  We investigate
the physical mechanisms responsible for the change in slope of this
relation, by comparing data with the results of the semi-analytic
model of galaxy formation {\morgana}, which already predicted such a
break in its original formulation. We find that the change of slope is
mostly induced by effective stellar feedback in star-forming
bulges. The shape of the relation is instead quite insensitive to
other physical mechanisms connected to BH accretion such as disc
instabilities, galaxy mergers, Active Galactic Nucleus (AGN) feedback,
or even the exact modelling of accretion onto the BH, direct or
through a reservoir of low angular momentum gas. Our results support a
scenario where most stars form in the disc component of galaxies and
are carried to bulges through mergers and disc instabilities, while
accretion onto BHs is connected to star formation in the spheroidal
component. Therefore, a model of stellar feedback that produces
stronger outflows in star-forming bulges than in discs will naturally
produce a break in the scaling relation. Our results point to a form
of co-evolution especially at lower masses, below the putative break,
mainly driven by stellar feedback rather than AGN feedback.
\end{abstract}


\section{Introduction}\label{sec:intro}

Since 1998, observations have shown the existence of more or less
tight correlations between the mass of the super-massive Black Holes
(BHs), hosted at the centre of local galaxies, and several properties
of their spheroidal component like luminosity
\citep[e.g.][]{Magorrian98, MarconiHunt03, Laesker14}, stellar mass
\citep[e.g.][]{HaringRix04, Sani11}, central velocity dispersion
\citep[e.g.][]{MerrittFerrarese01, Tremaine02, Shankar09} or radial
concentrate of stars \citep[e.g.][]{Graham01, Savorgnan13}.

The discovery of these scaling relations has been the starting point
for a re-evaluation of the role of accretion of gas into BHs and their
energetic feedback due to Active Galactic Nucleus (AGN) activity, in
the context of galaxy evolution.  Starting from this evidence, a
number of theoretical studies \citep[see e.g.][]{KauffmannHaehnelt00,
  Monaco00, Granato04, Hopkins06g, Fontanot06, Fanidakis12,
  Hirschmann12, Menci14} have investigated the physical mechanisms
responsible for these tight relations. Most of these studies have been
developed in the framework of ``co-evolution'' of BH growth and host
galaxies, where either the modulation of accretion or AGN feedback are
responsible for the correlations.  Other authors
\citep[e.g.][]{Peng07, JahnkeMaccio11} have instead pointed out that a
linear relation between bulge and BH masses is not necessarily a sign
of co-evolution, but is also comparable with a simple
``co-habitation'' model where correlations are mainly driven by galaxy
mergers, via the central limit theorem.

In the last years, new observations have revisited and sometimes
modified these scaling relations. In particular, recent updates have
been published pointing to higher normalizations in the BH mass-bulge
stellar mass \citep[e.g.,][]{Graham12, KormendyHo13} and BH
mass-central velocity dispersion relations \citep[e.g.,][]{Graham15}.
This in turn could imply significantly larger total mass densities for
the super-massive BHs in the local Universe \citep[e.g.,][]{Novak13,
  Shankar13, Comastri15}. These revisions are due to a number of
improvements both in the observational data, thanks to adaptive optics
and integral-field spectroscopy, and in the modeling of star
kinematics. Bulge masses are intrinsically more difficult to
determine, with respect to other bulge properties like luminosity or
velocity dispersion, due to the uncertainties in mass-to-light ratio
(which stem from the uncertainties in the initial stellar mass
function, the modelling of stellar evolution and the age-metallicity
degeneracy). A breakthrough in this respect came from the study of
galaxy kinematics in SAURON\footnote{Spectrographic Area Unit for
  Research on Optical Nebulae} and ATLAS-3D surveys
\citep{Cappellari06, Cappellari13}, which provided calibrated
relations between mass-to-light ratios and, e.g., velocity dispersion
for a representative sample of local galaxies. All these developments
led to a significant recalibration of the overall $M_{\rm BH}/M_{\rm
  sph}$ ratio, from $\sim 0.1$ percent \citep[e.g.][]{Sani11} to
$0.49$ percent \citep[e.g][]{GrahamScott13}.

Moreover, some evidence has also been accumulating suggesting a
scenario where different galaxy populations may follow individual (and
different) relations. For example, pseudo-bulges, usually defined as
spheroids characterized by disk-like exponential profiles and/or
rotational kinematics, were reported to systematically lie below the
main relation defined by classical bulges and ellipticals \citep[see,
  e.g., the discussion in][]{Graham08, Hu08, Shankar12, KormendyHo13},
but recent studies do not confirm these findings (Savorgnan et al., in
preparation). The power-law, almost linear relation (in logarithmic
space) between BH mass and either bulge stellar mass or central
velocity dispersion has been often used as a constraint for models of
the joint evolution of galaxies and AGNs. Indeed, although theoretical
models may trace different evolutionary paths for different galaxy
populations \citep[see, e.g.,][]{Lamastra10}, they typically predict a
local {\bbrel} compatible with a single power-law down to small mass
scales that are barely probed by observations
\citep[e.g.,][]{Marulli08, Menci08, Fanidakis12, Hirschmann12}. A
remarkable exception is provided by the MOdel for Rise of GAlaxies aNd
AGNs \citep[{\morgana},][]{Monaco07}, and presented in
\cite{Fontanot06}.  In that paper, the predicted {\bbrel} had a change
of slope at $M_{\rm BH} \sim 10^8$ M$_\odot$, with a steeper relation
at smaller masses.  This break had no observational support, but was
not incompatible with the sparse data available at that time.

In what follows, whenever we refer to the {\it Black Hole - Bulge
  relation} ({\bbrel}), we will specifically focus on the relation
between BH mass ($M_{\rm BH}$) and the stellar mass of the
spheroidal/bulge component of the host galaxy ($M_{\rm sph}$).

The overall shape of the {\bbrel} over a wide range in $M_{\rm sph}$
has been recently
revised by \citet{GrahamScott15}, who studied the {\bbrel} using data
from \citet{Scott13}, and included AGN data for which $M_{\rm sph}$
had been derived for the first time. These authors found that the
relation is linear for BH masses above $M_{\rm BH} \gtrsim 2\times
10^8\, M_\odot$, while it significantly steepens towards a quadratic
relation below this threshold. They argued that previous studies
missed the ``break'' in the {\bbrel} due to an insufficient sampling
at the low-mass end, mainly below $M_{\rm BH} \lesssim 10^7 M_\odot$.
It must be however stressed that
consensus on the statistical relevance of the break in {\bbrel} has
still to be reached. \citet{GrahamScott15} further revealed that the
most massive galaxies defining the linear part of the {\bbrel} are
mostly described by a core-S\'{e}rsic stellar profile, i.e a profile
with a deficit of light in central regions with respect to the
extrapolation of their outer profile, while the less massive ones are
better defined by a single S\'{e}rsic profile.

The existence of a quadratic, if not steeper, {\bbrel} has key
implications for the modeling of the joint evolution of BHs and their
host galaxies. In particular, it implies that the $M_{\rm BH}/M_{\rm
  sph}$ ratio is not constant, but it rather depends on the final
amount of stellar mass in the host galaxy, and/or its morphological
type. \citet{GrahamScott15} proposed a physical interpretation of
their results suggesting that the initial dissipative processes
controlled by the gas-rich, initial phases of galaxy-BH formation,
establish a quadratic relation.  Later, mostly dry (gas-poor) mergers
are responsible for the gradual build-up of the flatter, linear
portion of the {\bbrel}.

The aim of this work is to use {\morgana} to deepen our understanding
of the origin and evolution of the correlation between BHs and their
hosts, by searching for the key physical processes that have an
effective impact in shaping the {\bbrel} and its proposed break. More
specifically, we will explore the impact of key physical mechanisms
(like disc instabilities, galaxy mergers and stellar/AGN feedback) on
the observables of interest. {\morgana} relies on quite general
assumptions on how galaxy formation takes place inside dark matter
haloes, and as such it is representative of the whole class of
semi-analytic models (SAMs) of galaxy formation and evolution
\citep[see e.g][for comparison with other models in the
  literature]{Fontanot09b, Fontanot12a, Knebe15}.
  
In the following, we recalibrate the {\morgana} model to reproduce the
most recent determinations of the local {\bbrel} and BH mass function,
together with the AGN luminosity function along the cosmic history.
We confirm the predicted break in the {\bbrel}, and check that it is
compatible with the determinations of \citet{Graham12} and
\cite{Scott13}.  We investigate the origin of the break, and identify
stellar feedback in star-forming bulges as the mechanism responsible
for it. This points towards a co-evolution scenario in which stellar
feedback shapes the low-mass {\bbrel} in a way that is incompatible
with a mere co-habitation scenario.

The structure of this paper is as follows. We first summarise the key
aspects of the AGN modelling in {\morgana} in Sec.~\ref{sec:models},
and detail the different model variants considered in this work in
Sec.~\ref{sec:runs}. We then compare in Sec~\ref{sec:results} model
predictions with both the \citet{Scott13} data on the {\bbrel}, and
other physically linked observational constraints such as the
bolometric AGN luminosity function and the local BH mass function. We
then discuss the relative contribution of the different physical
mechanism included in our modeling, and present our conclusions in
Sec.~\ref{sec:final}.

\section{Semi-Analytic Model}\label{sec:models}

{\morgana}, first presented in \citet{Monaco07} and further updated in
\citet{LoFaro09}, relies on simplified modeling of those physical
processes that are believed to take place inside dark matter haloes
and drive the formation of galaxies. These can be broadly divided into
gravitational processes, such as stellar stripping, secular evolution,
galaxy interactions and mergers, and hydrodynamical/thermal processes,
such as gas cooling, star formation, BH accretion, stellar and AGN
feedback. The main strength of the semi-analytic approach is the
extensive use of approximate prescriptions, based on theoretical,
numerical, or observational results, to follow complex physical
processes in a simplified way.

We refer the reader to the original papers and to our comparison work
\citep[e.g.][]{Fontanot09b,DeLucia10,DeLucia11, Fontanot13} for a
complete overview of the model; in the next section we will mainly
focus on a single aspect of {\morgana}, namely the modelling of BH
accretion and AGN feedback \citep[see][for a complete overview of this
  approach]{MonacoFontanot05, Fontanot06}.

\subsection{BH accretion}

The model for accretion of gas onto the BH follows \cite{Granato04},
and starts from the assumption that the main bottleneck is given by
the loss of angular momentum, necessary for the gas to flow onto a
putative accretion disc. As soon as a galaxy is formed, the model
assumes that it contains a seed BH of $10^3\ M_\odot$. The first step
in the loss of angular momentum is connected to the same processes
(mergers, disc instabilities) that bring gas to the spheroidal
component; this assumption links BH growth with bulge formation events
in the history of the model galaxy.  Further loss is triggered by
other physical processes (i.e. turbulence, magnetic fields or
radiation drag) that are related to star formation in the bulge
\citep{Umemura00}. This results in the building up of a reservoir of
low angular momentum gas with mass ${M}_{\rm RS}$.  Its evolution is
regulated by growth and loss rates ($\dot{M}^{+}_{\rm RS}$ and
$\dot{M}^{-}_{\rm RS}$ respectively). The growth rate is taken to be
proportional to some power of the bulge star formation rate
($\phi_{\rm B}$):

\begin{equation}
\dot{M}^{+}_{\rm RS} = f_{\rm BH} \phi_{\rm B} \left( \frac{\phi_{\rm
    B}}{100 M_\odot yr^{-1}} \right)^{\alpha-1},
\end{equation}\label{eq:agn1}

\noindent
where $\alpha=1$ refers to the \cite{Umemura00} relation used in
\citet{Granato04}, while $\alpha=2$ corresponds to a model where the
loss of angular momentum is due by cloud encounters
\citep{Cattaneo05}. $f_{\rm BH}$ is a free parameter, whose value
affects the normalization of the final {\bbrel}, as it regulates the
total amount of cold gas flowing into the reservoir during galaxy
evolution. The gas in the reservoir then accretes onto the BH at a
rate regulated by the viscosity of the accretion disc
\citep{Granato04}:

\begin{equation}
\dot{M}^{-}_{\rm RS} = \dot{M}_{\rm BH} = 0.001 \frac{\sigma_B^3}{G} \left( \frac{M_{\rm
    RS}}{M_{\rm BH}} \right)^{3/2} \left( 1+\frac{M_{\rm BH}}{M_{\rm
    RS}} \right)^{1/2}
\end{equation}\label{eq:agn2}

\noindent
where $\sigma_B$ is the 1D velocity dispersion of the bulge and
$M_{\rm BH}$ is the BH mass. This term is also capped at the Eddington
limit.

A key role in the feeding of the central BH is played by the actual
amount of cold gas available in the spheroidal component as a
consequence of bulge formation events. A comprehensive discussion of
the relative importance of the different physical mechanisms
responsible for the building up of spheroidal components in galaxies
has been presented in \citet{DeLucia11}.

Here we simply recall that in {\morgana} galaxy mergers are
distinguished in major (baryonic merger ratio larger than $0.3$) and
minor mergers. In the former case, the whole stellar and gaseous
contents of both colliding galaxies are transferred to the spheroidal
remnant, while in the latter the total baryonic mass of the satellite
is given to the bulge of the central galaxy.

Secular evolution is an additional mechanism responsible for
transporting material to the centre of galaxies. In {\morgana} we use
the standard \citet{Efstathiou82} criterion to define the stability of
disc structures against self gravity. However, whenever a disc is
found to be unstable, no consensus has been reached yet on how to
model the resulting instability. Different SAMs have made different
choices, from moving just enough material to restore stability
\citep[e.g.,][]{Guo11,Hirschmann12}, to completely destroying the
unstable disc \citep[e.g.,][]{Bower06,Fanidakis12}.  \citet{Menci14}
used an updated version of a model by \citet{HopkinsQuataert11},
calibrated on results from numerical simulations. In all cases,
whenever an unstable disc contains cold gas, a fraction of it will be
made available to accrete onto the BH, powering an AGN.  In
{\morgana}, we assume that, when the instability criterion is met, a
fixed fraction ($f_{\rm DI}=0.5$) of the baryonic mass of an unstable
disc goes into the bulge component.  Cold gas flowing into the bulge
can then accrete onto the BH as described above. In what follows we
will discuss the impact of varying $f_{\rm DI}$ on our final results.

\subsection{AGN-triggered winds}

AGN activity can inject a relevant amount of energy into the
inter-stellar medium, favouring the triggering of a massive galactic
wind that halts the star formation episode ultimately responsible for
BH accretion. In the AGN feedback model included in {\morgana}, the
onset of such dramatic events is supposed to be due to the interplay
between stellar and AGN feedback, as described in
\cite{MonacoFontanot05}.  The triggering of a massive wind capable of
removing all cold gas from the host bulge, is subject to three
conditions.  The first condition is motivated by the theoretical
expectation that AGN radiation is able to evaporate some $50 M_\odot$
of cold gas for each $M_\odot$ of accreted mass.  Whenever the
evaporation rate overtakes the star formation rate, the inter-stellar
medium is expected to thicken and cause the percolation of all
Supernovae remnants into a galaxy-wide super-bubble, that can be
accelerated by radiation pressure from the AGN.  The second condition
requires that energy needed to sweep out all the inter-stellar medium
of the host bulge is not greater than the available energy from the
AGN.  The third condition is that the BH is accreting in a radiatively
efficient way, i.e. more than one percent of its Eddington limit.

In \citet{Fontanot06}, we showed that AGN-triggered winds are a key
ingredient to reproduce the evolution of the AGN luminosity function
and the space density of bright quasars in {\morgana}: therefore, in
this paper we will consider models including this mechanism. In
particular, we will refer to the {\morgana} version labelled ``dry
winds'', which implies using a high value of $f_{\rm BH}$ with winds
actually limiting the BH masses. It is worth noting, that
AGN-triggered winds are not necessary to reproduce the {\bbrel} in
{\morgana}. In a model without winds, however, the {\bbrel} is mainly
determined by the mechanism that regulates the accretion onto the BH,
rather than by feedback-based self-regulation. We refer the reader to
\citet{Fontanot06} for more details, here we just stress that, within
{\morgana}, the AGN-triggered winds have a limited impact on the
shape of the {\bbrel} (their Fig.~6), although they can affect
its redshift evolution (their Fig.~10).

\subsection{Role of stellar feedback}

The stellar feedback prescriptions included in {\morgana} follow the
results from the analytic model of \citet{Monaco04}, which postulates
different regimes according to the density and vertical scalelength of
the galactic system and distinguishes ``thin'' systems (i.e discs)
from ``thick'' systems (i.e. bulges and spheroids). In
\citet{Fontanot06}, we showed that stellar feedback has an overall
modest impact on the evolution of the AGN population, with the
relevant exception of \emph{kinetic} feedback in thick systems.
  
Kinetic stellar feedback is modelled by estimating the level of
turbulence in a star-forming bulge, quantified as the gas velocity
dispersion $\sigma_{\rm cg}$, due to the injection of kinetic energy
from Supernovae and the dissipation of turbulence on a sound crossing
time scale.  It is easy to show that this quantity scales with the gas
consumption time-scale of gas $t_\star$ to the power $-1/3$, so the
velocity dispersion of gas in star-forming bulges can be modeled as
follows:

\begin{equation}
\sigma_{\rm cg} = \zeta_0 \left( \frac{t_{\star}}{1\ {\rm Gyr}} \right)^{-1/3} {\rm km/s}
\end{equation}\label{eq:agn4}

\noindent
where the parameter\footnote{This parameter was named $\sigma_0$ in
  \citet{Fontanot06}.}  $\zeta_0$ regulates the efficiency at which
Supernovae drive turbulence.  Assuming that cold gas clouds have a
Maxwellian distribution of velocities, cold gas is then ejected into
the halo at a rate proportional to the probability that the velocity
of a cloud overtakes the escape velocity from the bulge.  So, the
combination of a significant $\zeta_0$ and a lower $t_\star$ leads to
high values of $\sigma_{\rm cg}$ in bulges, driving massive outflows
in small bulges.

In thin systems, where Supernova energy can be easily blown out of the
system, the value of $\zeta_0$ is expected to be smaller.  In
\citet{Fontanot06} we showed that kinetic feedback mechanism in
star-forming bulges is a viable mechanism to explain the downsizing
behaviour of the AGN population, as high velocity dispersions of cold
gas leads to massive removal of cold gas and to the suppression of
low-luminosity AGN at moderate-to-high redshifts.

\begin{figure}
  \centerline{
    \includegraphics[width=9cm]{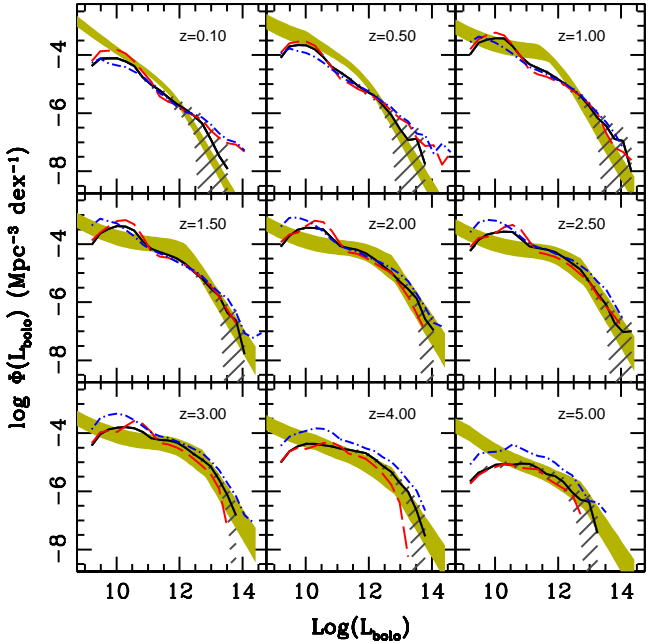} }
  \caption{Redshift evolution for the AGN bolometric luminosity
    function. Yellow areas represent an estimate for the bolometric
    luminosity function, obtained from the Hard X-ray LF
    \citep{Ueda14}. Black solid, dashed red and dot-dashed blue lines
    refer to the predictions of the reference, low-DI and high-DI runs
    respectively. Hatched areas show the 1-$\sigma$ variance
    associated with model predictions for the reference
    model.}\label{fig:bolo}
\end{figure}

\section{Runs}\label{sec:runs}
\begin{table}
  \caption{AGN feedback parameters for the main runs discussed in this
    work.}
  \label{tab:runs}
  \renewcommand{\footnoterule}{}
  \centering
  \begin{tabular}{cccc}
    \hline
     Model & $f_{\rm DI}$ & $\zeta_0$ [Km/s] & $f_{\rm BH}$ \\
    \hline
    Reference & 0.5 & 30 & 0.05 \\
    Low-DI    & 0.0 & 20 & 0.02 \\
    High-DI   & 1.0 & 20 & 0.003 \\
    \hline
  \end{tabular}
\end{table}

The reference AGN feedback model by \cite{Fontanot06} adopted here has
two main parameters that regulate the accretion rate of the cold gas
reservoir around the BH, namely $\zeta_0$, which regulates the amount
of stellar kinetic feedback, and $f_{\rm BH}$ which tunes the
normalization of the \bbrel. In addition, the overall radiative
efficiency $\eta$, which is the emitted luminosity in units of the
rest-mass energy, must be recalibrated when using a higher
normalization for the {\bbrel}.  This quantity is in fact broadly
proportional to the ratio between the integrated emissivity of AGN
over time and luminosity and the local mass density of BHs
\citep{Soltan82}. We thus decrease it from the original value of
$\eta=0.1$ adopted by \cite{Fontanot06} to $\eta=0.06$
\citep[see][]{ShankarWeinberg13}, and keep it always fixed to this
value in what follows.

All the runs presented in this work have been performed on the same cubic
box ($100 \, {\rm Mpc}$ side) obtained using the code {\sc pinocchio}
\citep{Monaco02, Monaco13} with $N=1000^3$ particles and assuming a
2006 concordance cosmology with parameters $\Omega_0=0.24$,
$\Omega_\Lambda=0.76$, $h=0.72$, $\sigma_8=0.8$, $n_{\rm sp}=0.96$. 

As the original {\morgana} runs were calibrated against the observed
{\bbrel} from \citet{MarconiHunt03}, we have recalibrated the model to
fit the relation proposed by \citet{GrahamScott15} ({\it reference}
run), and to reproduce the evolution of the AGN luminosity function.
Regarding the {\bbrel}, we decided to calibrate our models to
reproduce the most reliable BH mass estimates, in the bulge mass range
of $M_\star>10^{11} M_\odot$. The best-fit parameter values are
reported in Table~\ref{tab:runs}. Models tend to have marginally
steeper slopes in the high mass end of the {\bbrel}, so in general the
relation is underestimated at the bending point. It would be easy to
absorb this discrepancy using a different calibration choice; however,
we consider this conservative choice sufficient for the scope of this
paper.

\begin{figure*}
  \centerline{ \includegraphics[width=18cm]{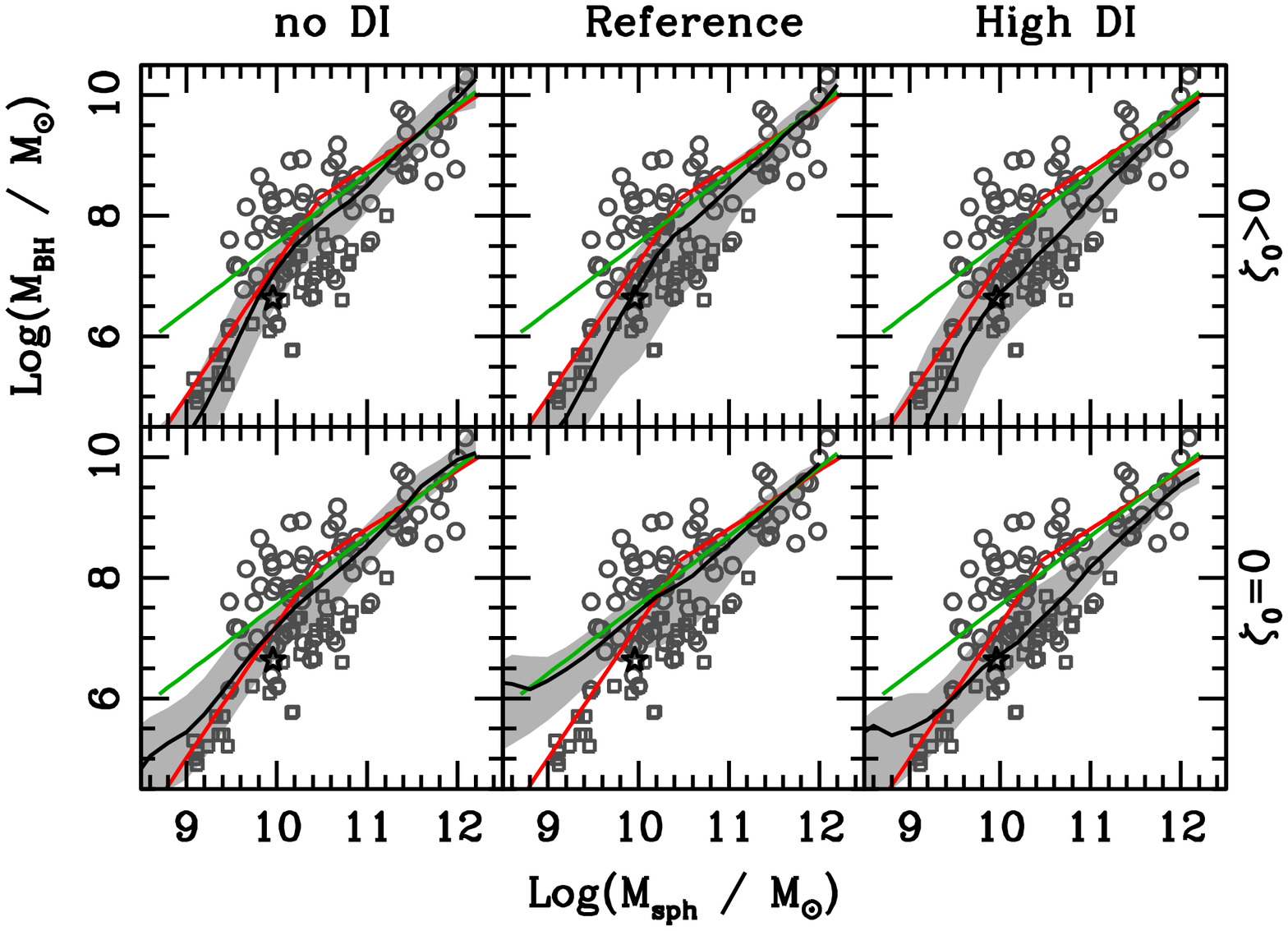} }
  \caption{{\bbrel} at $z=0$, for different models considered in this
    work. Red lines correspond to the best fits to the observations as
    in \citet[][]{Scott13}, while the green line refers to the best
    fit as in \citet{KormendyHo13}; open circles and squares show data
    from \citet{Scott13} and~\citet{GrahamScott15},
    respectively. Solid black lines show the median relation for model
    galaxies, while the grey area represents the $16\%$ and $84\%$
    percentiles.}\label{fig:bbrel}
\end{figure*}

\begin{figure*}
  \centerline{ \includegraphics[width=18cm]{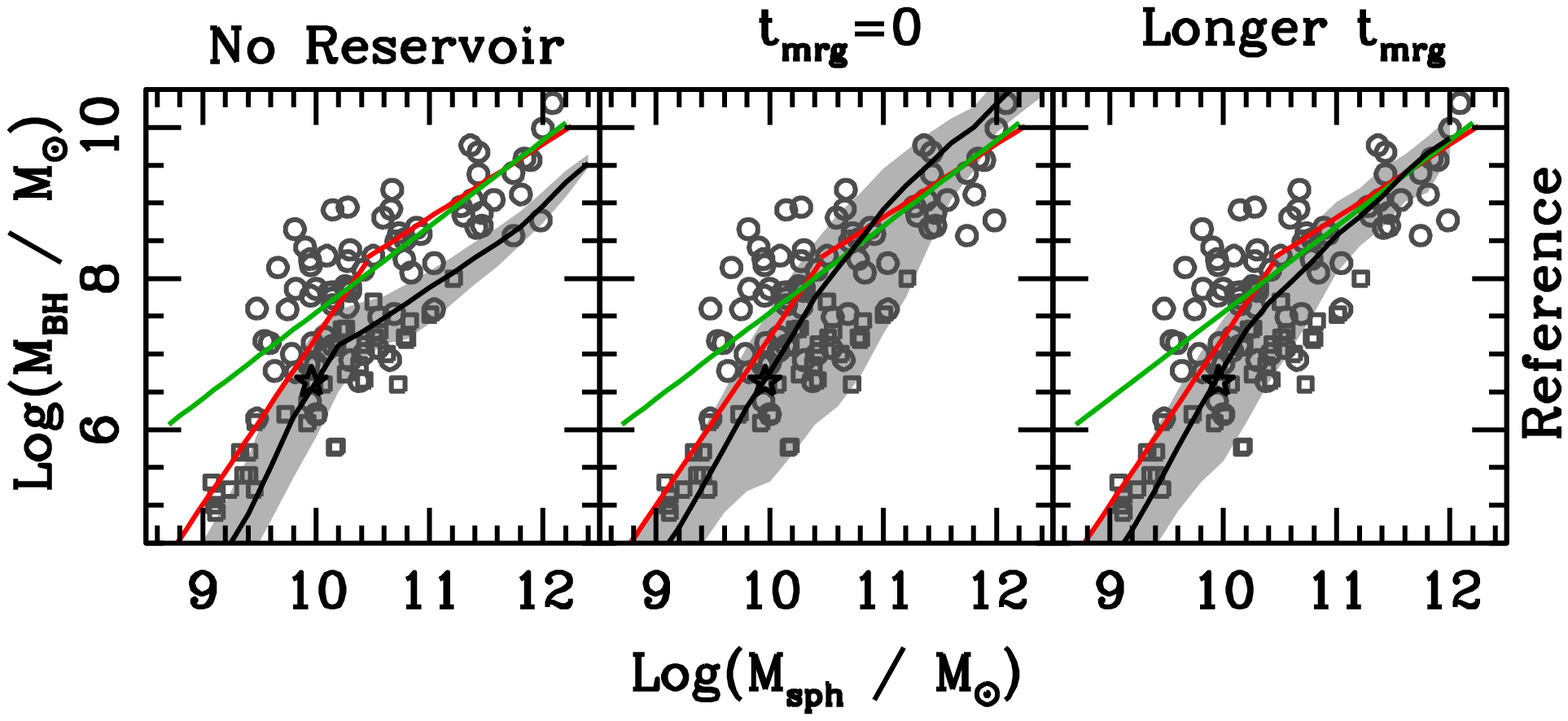} }
  \caption{{\bbrel} at $z=0$, for different models considered in this
    work. Lines, colours and shading as in
    Fig.~\ref{fig:bbrel}.}\label{fig:add_bbrel}
\end{figure*}

We then define some model variants, to explore the sensitivity of our
results to model assumptions. We first explore the impact of our
choice for the modelling of bar instabilities, by defining two
additional runs. In the first one, we switch off disc instabilities
($f_{\rm DI}=0$, {\it no-DI} run), while in the other, we assume that
the unstable disc is completely destroyed and the disc instability
results into a purely elliptical galaxy ($f_{\rm DI}=1$ {\it high-DI}
run).
These runs have also been recalibrated to fit the
\citet{GrahamScott15} {\bbrel}. Best-fit values for the relevant
parameters have been collected in Table~\ref{tab:runs}.

The AGN bolometric luminosity function used for calibration is
reported in Fig.~\ref{fig:bolo}. Here the reference run is shown as a
black solid line, while dashed red and dot-dashed blue lines refer to
the no-DI and high-DI runs respectively.  The observational estimate
of the bolometric luminosity function (yellow area) is obtained from
the Hard X-ray LF from \citet[yellow shaded area]{Ueda14}, using both
the \citet{Marconi04} bolometric correction and an estimate for the
fraction of Compton-thick objects by \cite{Ueda14}. All three runs
broadly reproduce the evolution of the bolometric AGN-LF over wide
ranges of redshifts and luminosities, although the high-DI model tends
to overproduce AGNs, especially faint ones, at high redshift. It is
worth noting that the new calibrations assume relatively low values
for $\zeta_0$ with respect to those quoted in
\citet{Fontanot06}. This implies that the impact of this parameter to
regulate the downsizing behaviour of the AGN population is reduced and
also runs with $\zeta_0=0$ (see below) provide a fair agreement with
the data.

For each of the three disc instability options we perform additional
{\morgana} runs by switching off the stellar kinetic feedback
($\zeta_0=0$).  These model variants have {\it not} been recalibrated
with respect to the parameters used in the $\zeta_0>0$ runs. A
further test has been performed by running the model without using the
cold, low angular momentum gas reservoir, i.e. assuming that a
(Eddington limited) fraction $\dot{M}^{+}_{\rm RS}$ of the cold gas in
the bulge is directly accreted by the BH.  Finally, we have studied
the role played by galaxy mergers on the {\bbrel}.

In {\morgana}, the merger times for satellite galaxies\footnote{Galaxy
  merger times are computed whenever a galaxy becomes a satellite,
  following the merger of its parent DM halo with a larger one.}
($t_{\rm mrg}$) is computed using an updated version of the fitting
formulae provided by \citet{Taffoni03}.  As shown in
\citet{DeLucia10}, over the relevant range of mass ratios, these
$t_{\rm mrg}$ are typically shorter than those estimated from other
authors, using different and more recent numerical simulations. We
then present two model variants obtained using different $t_{\rm mrg}$
definitions, namely the fitting formulae proposed by \citet[``longer
  $t_{\rm mrg}$'']{BoylanKolchin08} and the more extreme case of
instantaneous mergers ($t_{\rm mrg}=0$).  Also in these three
additional versions we kept the other parameters as in the $\zeta_0>0$
runs.

It is also worth stressing, that in all the models discussed in this
work we do change only the subset of parameters directly connected to
AGN modelling, while keeping all other parameters to the values
defined in \citet{LoFaro09}, with the exception of the recomputation
of the size of starbursts\footnote{The computation of gas consumption
  time-scale $t_\star$, based on the \cite{Kennicutt98} relation and
  thus on gas surface density, requires knowledge of the size of the
  star-forming region. This quantity was set equal to the bulge size
  in the original {\morgana}, while in \citet{LoFaro09} a more refined
  modeling was adopted, based on the idea that the level of turbulence
  should equate the velocity given by the bulge rotation curve
  computed at the starburst size. Nonetheless, this change caused a
  drastic steepening of the {\bbrel}, so we switched it off to achieve
  a good modelling of AGNs. The impact on other observables is modest,
  and can be absorbed by suitable retuning of other parameters.}.

This implies that, in general, the agreement of model predictions with
the calibration set (i.e. local stellar mass function, cosmic star
formation rates and morphological mix) may not necessarily be
optimal. We, however, explicitly tested that the deviations from the
reference models for non-AGN galaxy properties are small. For
instance, stellar mass functions are relatively weakly affected by the
exact AGN modelling adopted, once the model is calibrated to reproduce
the {\bbrel} and the AGN luminosity function, while the morphological
mix is more sensitive to the $\zeta_0$ parameter and to disc
instabilities. We checked that our reference run produces a plausible
morphological mix.

\section{Results}\label{sec:results}

\begin{figure}
  \centerline{
    \includegraphics[width=9cm]{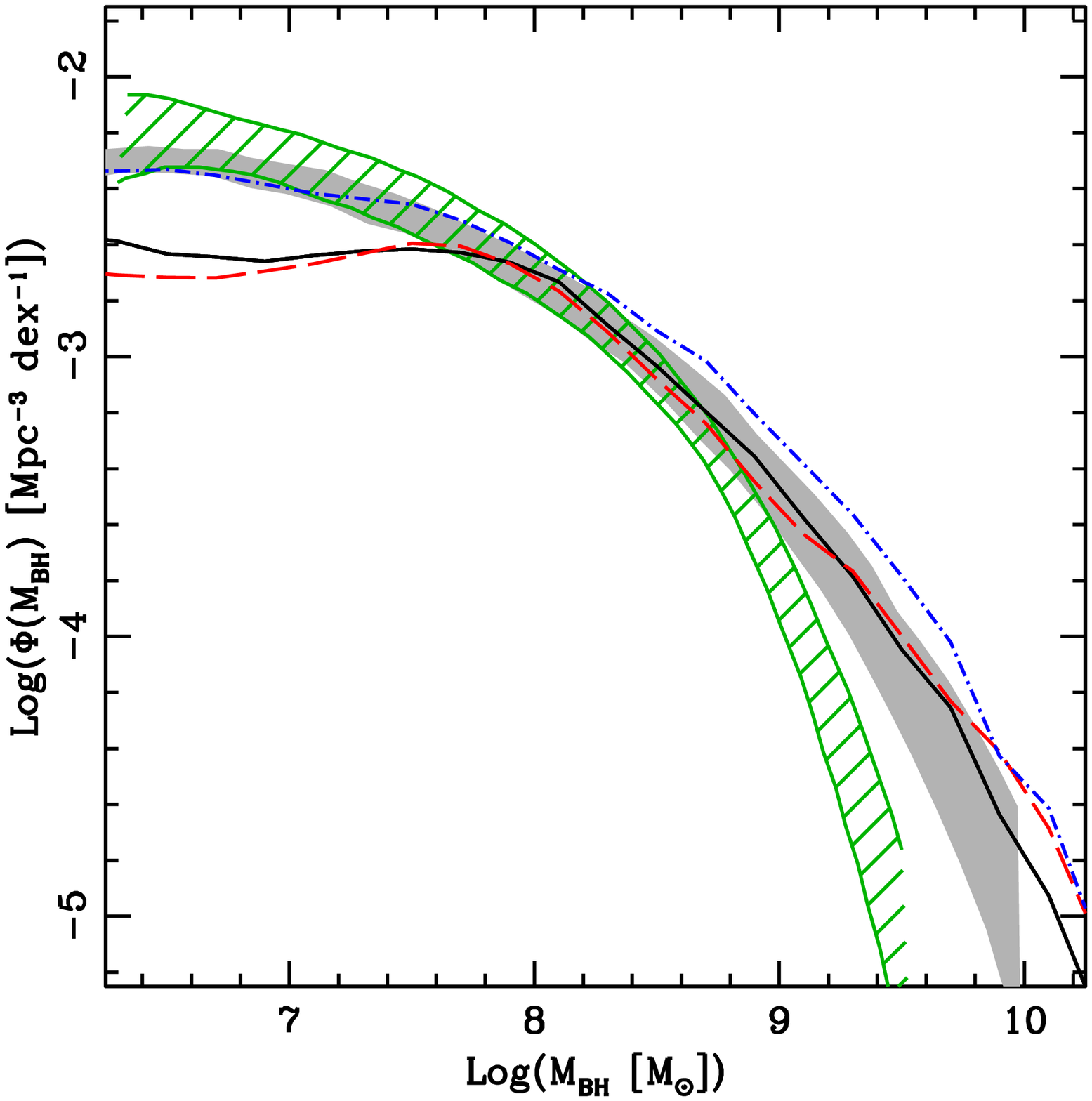} }
  \caption{BH mass function at $z=0$. Data from \citet[][grey
      area]{Shankar13} \citet[][green hatched
      area]{Shankar09}. Linetypes and colours as in
    Fig.~\ref{fig:bolo}.}\label{fig:mfbh}
\end{figure}

\begin{figure}
  \centerline{
    \includegraphics[width=9cm]{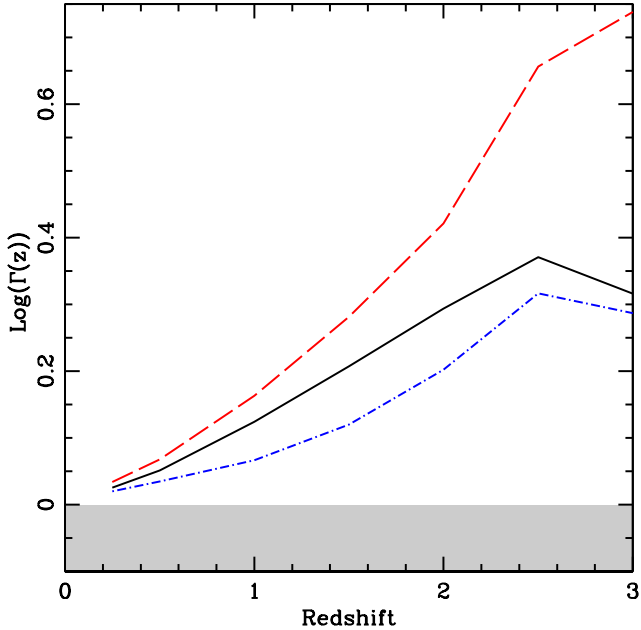} }
  \caption{Redshift evolution of the overall {\bbrel}. Line types and
    colours as in Fig.~\ref{fig:bolo}. See text for more details on
    the definition of the $\Gamma$ parameter.}\label{fig:bbevo}
\end{figure}
The $z=0$ {\bbrel} for the runs defined in the previous section are
shown in Fig.~\ref{fig:bbrel}. In all panels, the black solid line is
the median relation for the model galaxy sample, while the grey area
represents the 16th percent and 84th percent percentiles of the
distribution.
%
%
The central panel in the upper row displays the {\bbrel} in our
reference run, compared to the observational relations proposed by
\citet[green solid line]{KormendyHo13} and \citet[red solid
  line]{GrahamScott15}. A steepening of the {\bbrel} is evident for
$M_\star \lesssim 10^{10} M_\odot$ galaxies, resulting in a break of
the relation similar to the results of \citet{GrahamScott15} (red
lines represent their best fit to power-law-S\'ersic and core-S\'ersic
samples).  Moreover, the scatter in the model is broadly compatible
with that in the data. The \citet{Scott13} datapoints suggest a larger
scatter than the predicted one. Note, however, that our model does not
include observational errors on BH and stellar masses. Also, the
intrinsic scatter reported by \cite{KormendyHo13} for the subsample
inclusive of only the most accurate BH mass estimates, amounts to
$\sim 0.3$ dex, even smaller than what predicted by our models for the
total sample.

We now discuss the implications for the {\bbrel} when varying one or
more of the input assumptions in our reference model. We first
consider the effect of changing our modelling for disc instabilities
(Fig.~\ref{fig:bbrel}, upper row). We remind that the no-DI and
high-DI runs require recalibration of the $\zeta_0$ and $f_{\rm BH}$
parameters to reproduce the observed {\bbrel}, due to the different
predicted amounts of cold gas available for BH accretion and AGN
feedback in the models. For all the runs we find a clear break in the
{\bbrel}, thus concluding that its presence does not depend on the
exact modelling of disk instabilities. The scatter, compared to the
reference run, decreases in the no-DI case, but remains constant in
the other case.  This is at variance with the results presented,
i.e. by \citet{Menci14}. We will deepen into this point in
Section~\ref{sec:final}.

We then turn to the predicted BH mass function (Fig.~\ref{fig:mfbh}),
and compare it with the empirical estimates by \citet{Shankar13},
based on the convolution between the galaxy velocity dispersion
function from \citet{Bernardi10}, and the BH mass-velocity dispersion
relation by \citet{McConnellMa13}, which is the most appropriate for
our renormalized-high BH-stellar mass relation. For completeness, we
also report the estimates by \citet{Shankar09}, which were anyway
based on previous (lower) normalization of the BH scaling
relations. All our runs show a reasonable agreement with the most
recent estimate at the high-mass end. At low BH masses the models with
no or moderate DI tend to underpredict the mass function, which was
anyway estimated assuming a strictly linear power-law scaling
relations. The model with strong {\it DI} provides a better fit, at
the cost of an overprediction at large masses\footnote{This trend
  cannot be predicted by the results of Fig.~\ref{fig:bbrel}, because
  it is driven by the larger mass in bulges that is obtained in this
  model. A full recalibration of the model, including also galaxy
  observables, may absorb this difference, although such a strong role
  of disc instabilities may easily result in an overproduction of
  bulges.}. We caution that the comparison with the local BHMFs is
simply provided as a broad consistency check. As discussed by several
authors \citep[see e.g.][]{Tundo07, Shankar12}, these estimates are
still affected by systematics on the exact normalization of the input
scaling relations and on the actual contributions of low-mass black
holes, which can be connected either to possible breaks or other
deviations (like, e.g., those proposed for the ``pseudo''-bulge
population) from the scaling relations defined by more massive black
holes.

The results shown in Fig.~\ref{fig:bolo} and~\ref{fig:mfbh} confirm
our previous conclusion that models with different strength of disk
instability provide very similar descriptions of the evolution of the
AGN population.

We then use our additional model variants to understand which physical
mechanism implemented in {\morgana} is the main driver for the break
in the theoretical {\bbrel}. In the lower row of Fig.~\ref{fig:bbrel}
we show {\morgana} runs where we switched off kinetic stellar feedback
in bulges (i.e. $\zeta_0=0$). All the resulting relations are power
laws over the whole $M_{\rm sph}$ range, and follow a relation with a
slope compatible to, or slightly steeper than, the
\citet{KormendyHo13} best-fit relation (green line), with no apparent
bend at the low-mass end. We thus conclude that the treatment of
stellar feedback is the main responsible for the different behaviours
at the high- and low-mass end of the {\bbrel}.

This is confirmed in Fig.~\ref{fig:add_bbrel}, where we report the
{\bbrel} predicted by the variants of the reference model without the
reservoir of low angular momentum gas, and with varied merging times.
We recall that these additional variants do use the same calibration
as the reference run, so the normalization of the {\bbrel} is not
guaranteed to be reproduced. All relations show the same steepening at
low masses. Moreover, the typical $M_{\rm sph}$ mass scale
corresponding to the resulting break does not change among the various
runs (with a possible exception of the zero merger time run),
confirming that stellar feedback is setting the mass scale of the
change of slope. Changing the merger timescale in {\morgana} has only
a limited effect on the shape and normalization of the {\bbrel}, and
we checked that this is independent of the assumed disc instability
model.

We finally address the redshift evolution of the {\bbrel}. We checked
that all models predict the {\bbrel} to be already in place at high
redshift, with almost no evolution in the shape with respect to the
local relation.  It is worth stressing that the actual detectability
of a ``bended'' relation at higher redshifts is related to the
abundance of the massive galaxy population.

We find some evolution in the normalization of the {\bbrel}. We quantify
this evolutionary trend by means of the $\Gamma(z)$ parameter, defined
as:

\begin{equation}
\Gamma(z) = \sum_{M_{\rm sph}} \frac{M_{\rm BH} (z)}{M_{\rm BH}
  (z=0)} (M_{\rm sph})
\end{equation}

\noindent
In Fig.~\ref{fig:bbevo} we show the average evolution of $\Gamma$,
computed with $M_{\rm sph}>10^9 M_\odot$ model galaxies, using the
corresponding median relations shown in Fig.~\ref{fig:bbrel} as $z=0$
reference.  The evolution up to $z=2$ is similar for all models, but
disc instabilities lead to a flattening of $\Gamma$ at high redshift.

\section{Discussion \& Conclusions}\label{sec:final}
\begin{figure*}
  \centerline{
    \includegraphics[width=18cm]{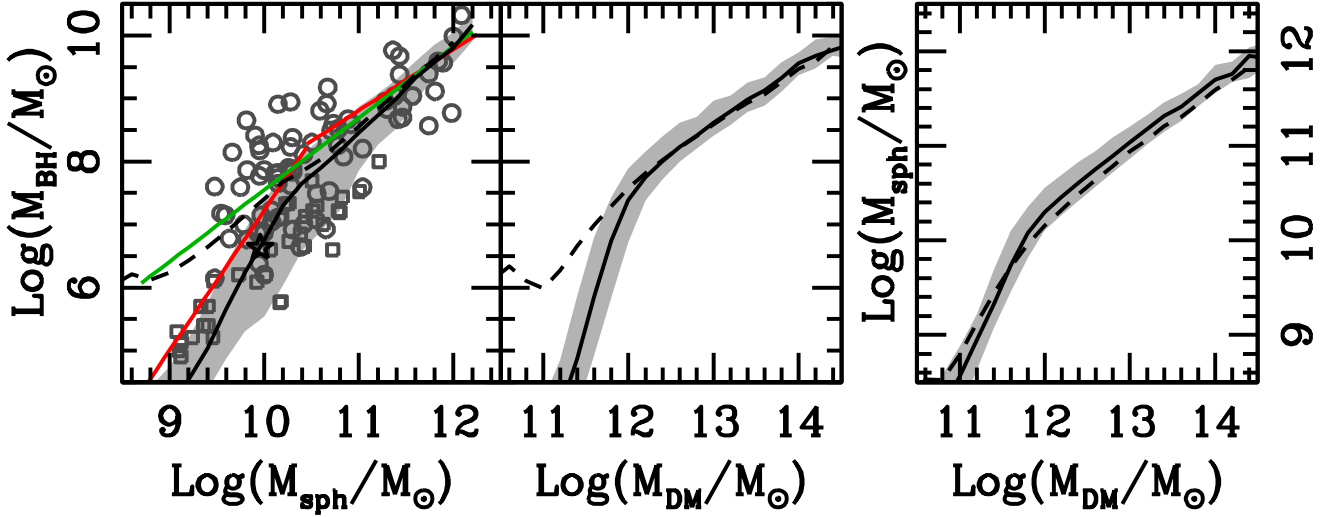} }
  \caption{Additional black hole - spheroid scaling relations for at
    $z=0$, for the different models considered in this work. {\it Left
      panel:} {\bbrel} relation (data as in
    Fig.~\ref{fig:bbrel}). {\it Middle panel:} $M_{\rm BH}$-$M_{\rm
      DM}$ relation. {\it Right panel:} $M_{\rm sph}$-$M_{\rm DM}$
    relation. In all panels, solid lines refer to the median relations
    in reference model (shaded areas representing the the $5\%$ and
    $95\%$ percentiles), while the dashed line to the corresponding
    $\zeta_0=0$ runs.}\label{fig:dmp}
\end{figure*}

Recent re-calibrations of the locally measured {\bbrel} suggest its
normalization to be a factor of $\sim 2$ higher than previously
inferred \citep[e.g.,][]{GrahamScott13, KormendyHo13}. Some authors
\citep{Scott13,GrahamScott15} have found a strong, quadratic
steepening of the {\bbrel} below $M_{\rm BH} \lesssim 2\times 10^8\,
M_\odot$.  The {\morgana} model made a prediction of such a steepening
at low masses in 2006 \citep{Fontanot06}.  While consensus on this
break in the relation is still to be achieved, in this paper we have
further investigated the possible physical causes of such a
steepening.  To this aim, after recalibrating the predicted {\bbrel}
to take account of the higher normalization, we explored the impact of
varying the strength of the disc instabilities, galaxy mergers and
stellar feedback in star-forming spheroids.

We explicitly tested that the strength of the disc instability, i.e.,
the amount of baryonic material transferred from the unstable disc to
the central spheroidal component, does not affect the shape of the
{\bbrel}, although it may change its normalization.  We also showed
that neither the assumed modelling of a gas reservoir around the BH,
nor the timing of galaxy mergers have a relevant effect on the shape
of the {\bbrel} in {\morgana}, while they have an influence on its
normalization and scatter.

In more detail, we find that the scatter in the {\bbrel}, as predicted
by {\morgana} does not depend on the modeling of disc instabilities,
i.e., on the relative importance of this physical mechanism in feeding
the central BH. This result may be reconciled with the results from
other SAMs \citep[see e.g][]{Menci14}, by considering the role of the
gas reservoir in our model. As a matter of fact, comparing the runs
assuming direct (Eddington limited) accretion onto the central BH,
i.e., switching off the modeling of the gas reservoir, we confirm a
decrease (of a factor about 1.5) in the scatter at the high-mass end
of the relation moving from the reference to the high-DI runs. This is
compatible with the idea that the delayed accretion from the reservoir
(with respect to the time when cold gas looses most of its angular
momentum) adds another time scale to the mechanism of BH accretion
that couples to disc instabilities. This additional time scale
dominates the scatter in the \bbrel.

The physical process responsible for the break in the {\bbrel} in
{\morgana} is stellar feedback in star-forming bulges.  Its role can
be illustrated as follows. Both bulges and BHs acquire stars from
merging and local star formation/accretion.  While most stars are
found in bulges at $z=0$ \citep[see e.g.][]{Gadotti09}, star formation
mostly takes place in the disc components of galaxies, with a minor
contribution from starbursts.  This is true in {\morgana}
\citep{Monaco07}, but observations suggest a similar trend at
$z\sim2$, with normal star-forming galaxies being predominantly
rotating discs \citep{ForsterSchreiber06} and starbursts lying above
the main sequence of starforming galaxies, contributing only $\sim10$
per cent to the total SFR density
\citep{Daddi10,Rodighiero11,Sargent12}.  As a consequence, most stars
in bulges were formed in discs and were carried into the spheroidal
component by mergers and disc instabilities.  Conversely, most mass in
the BHs is brought by accretion and not by mergers
\citep[e.g.][]{Salucci99, MarconiHunt03}.  As long as BH accretion is
related to star formation in bulges, a selective suppression of this
star formation in small bulges will decrease small BH masses but will
not influence much the stellar mass of bulges.

This interpretation of the break {\bbrel} differs from the
\citet{GrahamScott15}, and we propose it as an alternative explanation
of the steepening of the {\bbrel} at small masses.  In particular, our
model does not produce two clearly different regimes as a function of
$M_{\rm sph}$. Star formation\footnote{In detail, star formation is
  still appreciable for massive galaxies in {\morgana}, due to the
  inefficient quenching of cooling flow in massive haloes for an AGN
  modelling, which does not include hot gas accretion
  \citep{Fontanot07b}.}, BH accretion and mergers (both dry and wet)
happen at all mass scales, although the relative importance of each
physical process depends on the total mass of the host galaxy
\citep{DeLucia11,Shankar13SAM}.  We also showed that the the shape of
the {\bbrel} depends weakly on the choice of merger timescales. In
particular, a significant shortening of the merger timescales (i.e.,
assuming instantaneous galaxy mergers) increases the mass of the most
massive BHs by a factor of up to $\sim 2$ and changes the position of
the break, but the overall {\bbrel} still shows a bend at the low-mass
end.

We also stress that the stellar feedback mechanism we propose is an
example of a possible self-regulation of BHs and bulges, where
effective stellar feedback leads to starvation of small BHs.  In this
sense, a steepening of the {\bbrel} can be seen as an argument in
favour of the ``co-evolution'' scenario as opposed to the
``co-habitation'' one, where a linear relation driven by mergers is
the most natural expectation.  This self-regulation is not due to AGN
feedback, although, as commented in Section~\ref{sec:models}, the
final BH mass is self-regulated by AGN feedback. Indeed, of the three
models presented in the original \cite{Fontanot06} paper, two were
based on a self regulation of BH masses due to AGN feedback, but all
three showed a break in the {\bbrel}.

Previous work already showed that non-trivial AGN feedback schemes,
whose efficiency varies as a function of halo mass, can create a break
in scaling relations, including the {\bbrel} \citep{Cirasuolo05,
  Shankar06}, though possibly milder than the one emphasised by
\citet{GrahamScott15}.  

It is at this point natural to ask whether a break in the BH-stellar
mass relation corresponds to a break in other scaling relations,
especially between $M_{\rm BH}$ and velocity dispersion $\sigma$,
usually measured to be even tighter. \citet{GrahamScott15} have shown
that no apparent break is evident in their data when plotting BH mass
versus velocity dispersion, while the Supernova feedback model
implemented by \citet{Cirasuolo05} predicted a clear break also in
this relation. An estimate of the central velocity dispersions of
bulges from {\morgana} would require to make several additional
assumptions on the exact light and mass profiles of the stellar, gas,
and dark matter components. Instead, we choose to consider the
predicted relation between $M_{\rm BH}$ and halo mass $M_{\rm DM}$:
this is shown in Fig.~\ref{fig:dmp}, where the left panel is a
replication of the reference model relation in Fig.~\ref{fig:bbrel}
(continuous line) and, overplotted, the median relation for the model
with $\zeta_0=0$ (dashed line). The other two panels of this figure
show the relation between BH mass and halo mass (middle), and between
halo mass and stellar mass of the spheroidal component (right). A
break in the $M_{\rm BH}$-$M_{\rm DM}$ relation is evident in our
reference model, while it is nearly absent in the model without
kinetic feedback (dashed lines). It is interesting to notice that the
shape of the $M_{\rm sph}$-$M_{\rm DM}$ relation is almost unaffected
by this mechanisms, confirming that this modelling of feedback in
star-forming bulges affects BH masses much more than the spheroidal
component of the galaxy.

It is worth stressing that it is not straightforward to infer
conclusions on the shape of the $M_{\rm BH}$-$\sigma$ relation from
the results shown in Fig.~\ref{fig:dmp}. While we expect $M_{\rm DM}$
to correlate with virial velocity, the relation between the latter
quantity and velocity dispersion is still uncertain for large galaxy
samples. Several groups \citep[see e.g.][]{FerrareseMerritt00, Baes03}
suggested that $\sigma$ correlates with the circular velocity as
measured at the outermost radii, while other authors \citep{Ho07}
showed that such a correlation has a significant scatter and varies
systematically as a function of galaxy properties. Therefore, any
further conclusion on the $M_{\rm BH}$-$\sigma$ relation critically
depends on the assumed relation between velocity dispersion and
circular velocity, e.g. if we assume a linear correlation, we then
expect a break also in the this relation.

Of course, our modelling of stellar and AGN feedback still represents
an idealized approach and we argue that observations of spatially
resolved galaxies (like e.g. CALIFA\footnote{Calar Alto Legacy
  Integral Field Area Survey}, \citealt{Sanchez12}) will provide a
better comprehension of the interplay between these two fundamental
physical mechanisms.

\section*{Acknowledgements}

We thanks Alister Graham, Alessandro Marconi and Giulia Savorgnan for
interesting discussions and useful suggestions. FF acknowledges
financial support from the grants PRIN INAF 2010 ``From the dawn of
galaxy formation'' and PRIN MIUR 2012 ``The Intergalactic Medium as a
probe of the growth of cosmic structures''.  PM acknowledges support
from PRIN INAF 2014 ``Glittering kaleidoscopes in 9the sky: the
multifaceted nature and role of Galaxy Clusters''.

\bibliographystyle{mn2e}
\bibliography{fontanot}

\end{document}